\documentclass[reprint,aps,prb,floatfix]{revtex4-1}

\usepackage{amsfonts, amsmath, amsthm, amssymb} 
\usepackage{wrapfig}
\usepackage{hyperref}
\usepackage{graphicx}

\newcommand{\Eqref}[1]{Eq.~(\ref{#1})}
\newcommand{\Fref}[1]{Fig.~\ref{#1}}

\begin{document}

\title[Generation of Electric Oscillations by Negative Differential Conductivity]
{Computer Simulation of Electronic Device for Generation of Electric Oscillations by Negative Differential Conductivity of Supercooled Nanostructured Superconductors in Electric Field}

\author{Todor M. Mishonov, Victor I. Danchev}

\affiliation{Department of Theoretical Physics, Faculty of Physics, Sofia University St. Kliment Ohridski, 5 James Bourchier Blvd., BG-1164 Sofia, Bulgaria}
\email{mishonov@bgphysics.eu}

\author{Ioulia~Chikina}
\affiliation{$^2$LIONS, NIMBE, CEA, CNRS, Universit\'e Paris-Saclay, CEA Saclay 91191 Gif sur Yvette Cedex, France}

\author{Albert M. Varonov}
\affiliation{Faculty of Physics, Sofia University St. Kliment Ohridski, 5 James Bourchier Blvd., BG-1164 Sofia, Bulgaria}

\date{28 May 2019}

\begin{abstract}
We use the formerly derived explicit analytical expressions for the conductivity of nanostructured superconductors supercooled below the critical temperature in electric field.
Computer simulations reveal that the negative differential conductivity region of the current-voltage characteristic leads to excitation of electric oscillations.
We simulate a circuit with distributed elements as a first step to obtain gigahertz oscillations.
This gives a hint that a hybrid device of nanostructured superconductors will work in terahertz frequencies.
If the projected setup is successful, we consider the possibility for it to be put in a nanosatellite (such as a CubeSat).
A study of high temperature superconductor layers in space vacuum and radiation would be an important technological task.
The interface thermal boundary resistance is taken into account in the numerical illustrations for the work of the device.
\end{abstract}

\maketitle

\section{Introduction}
The purpose of the present work is to present a numerical simulation which shows the generation of high-frequency oscillations created by a thin superconducting film supercooled bellow the critical temperature $T_c$ in external electric field $E$. The idea was first proposed more than 10 years ago \cite{TerraHz} and now we are giving details on how to start the experimental work using radio frequency oscillations with distributed elements (resistors, capacitors and inductors). The main idea is to insert a supercooled superconductor with negative differential conductivity as an active element in a resonance circuit. The simulation presented in this work aims to trigger the analogous development of terraherz region where the resonance circuit will be implemented on the same hybrid nanodevice which will emit coherent terraherz waves.

\section{Fluctuation conductivity in electric field}
Our starting point is the result for the two-dimensional fluctuational current density $j_\mathrm{_{2D}}$ in external electric field\cite{Kinetics,Fluctuation,Varlamov,Mishonov}
\begin{equation} 
j_\mathrm{_{2D}}=\sigma_0\left[ S_n+ S(\epsilon, \beta)\right] E,
\label{VAchar}
\end{equation}
where $\beta$ is the dimensionless electric field and $\epsilon$ is the reduced temperature
\begin{equation}
\beta=\frac{\pi e \xi E}{16 k_\mathrm{_B} T_c} 
=\frac{e\xi E}{\hbar/\tau_0}, \qquad \epsilon = \ln \frac{T}{T_c} \approx \frac{T-T_c}{T_c} 
\end{equation}
defined by the coherence length $\xi$
\begin{equation}
\left. -T_c \frac{\mathrm{d} B_\mathrm{c2}}{\mathrm{d}T} \right \vert_{T_c}
= \frac{\Phi_0}{2 \pi \xi^2}, \qquad \Phi_0 = \frac{\pi \hbar}{| e |},
\end{equation}
the critical temperature $T_c$ or a time constant $\tau_0$, related to the relaxation of the superconducting order parameter
\begin{equation}
\hbar/\tau_0=\frac{16  k_\mathrm{_B} T_c}{\pi}.
\end{equation}
The constant $\sigma_0$ with dimension of conductivity
\begin{equation}
\sigma_0=\frac{e^2}{16 \hbar}
\end{equation}
represents the Aslamazov-Larkin conductivity above $T_c$ for evanescent electric field $E \rightarrow 0$
\begin{equation}
\sigma_{\mathrm{AL}}=\frac{e^2}{16 \hbar \epsilon}, \qquad \epsilon > 0.
\end{equation}
The normal conductivity $\sigma_n$ is parameterized by the dimensionless $S_n=\sigma_n/ \sigma_0$.

\begin{figure}[h]
\includegraphics[width=1\linewidth]{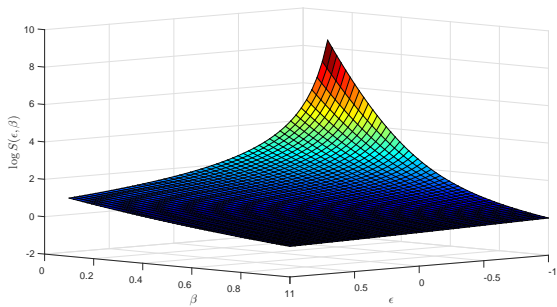}
\caption{The logarithm of the dimensionless conductivity $S(\epsilon,\beta)$ as a function of reduced temperature $\epsilon$ and dimensionless electric field $\beta$ \Eqref{eqS}.}
\label{Fig1}
\end{figure}

The dimensionless function $S(\epsilon, \beta)$ of reduced temperature $\epsilon$ and electric field $\beta$ 
\begin{equation} \label{eqS}
S(\epsilon, \beta)=\int\limits_0^\infty \exp\left(-\epsilon v - \frac{\beta}{3} v^3 \right) \mathrm{d} v,
\end{equation}
describing the fluctuational conductivity is depicted in \Fref{Fig1}.
This formula for the current \Eqref{eqS} is a consequence of time-dependent Ginzburg-Landau equation, which is applicable below $T_c$ if the coherent superconducting order parameter is destroyed by constant electric field.
For $\epsilon=0$ the integral is elementary.
In other cases the substitution $u=|\epsilon|v$ reduces the integral to well-known special functions.

For low frequencies where the heat capacity of the superconductor is negligible, the temperature of the superconducting layer is determined by the Ohmic heating per unit area $E j_\mathrm{_{2D}}(E)$ and interface boundary resistance $\mathcal{R}_\theta$
\begin{equation}
E j_\mathrm{_{2D}}(E,T) =(T - T_\mathrm{sub})/\mathcal{R}_\theta, \qquad
\rho_\theta = \frac{16 k_\mathrm{_B}^2 T_c \mathcal{R}_\theta}{\pi^2 \hbar \xi^2},
\label{Kapitsa}
\end{equation}
where $\rho_\theta$ is a convenient dimensionless parameter useful for the further analysis and $T_\mathrm{sub}$ is the substrate temperature.
Until the critical temperature $T_c$ can be found by fitting the temperature dependence of the resistivity above $T_c$, the coherence length $\xi$ can be extracted without applying magnetic field $B$ from the non-linear fluctuational conductivity~\cite{Hurault} at $T_c$
\begin{equation}
j_\mathrm{_{2D}} = \dfrac{3^{1/3} \Gamma\left(\frac43\right)}{2^{4/3} \hbar}
\left(\frac{k_\mathrm{_B} T_c}{\pi e \xi} \right)^{2/3} E^{1/3} + \sigma_n E, \quad T=T_c.
\end{equation}

\begin{figure}[h]
\includegraphics[width=1\linewidth]{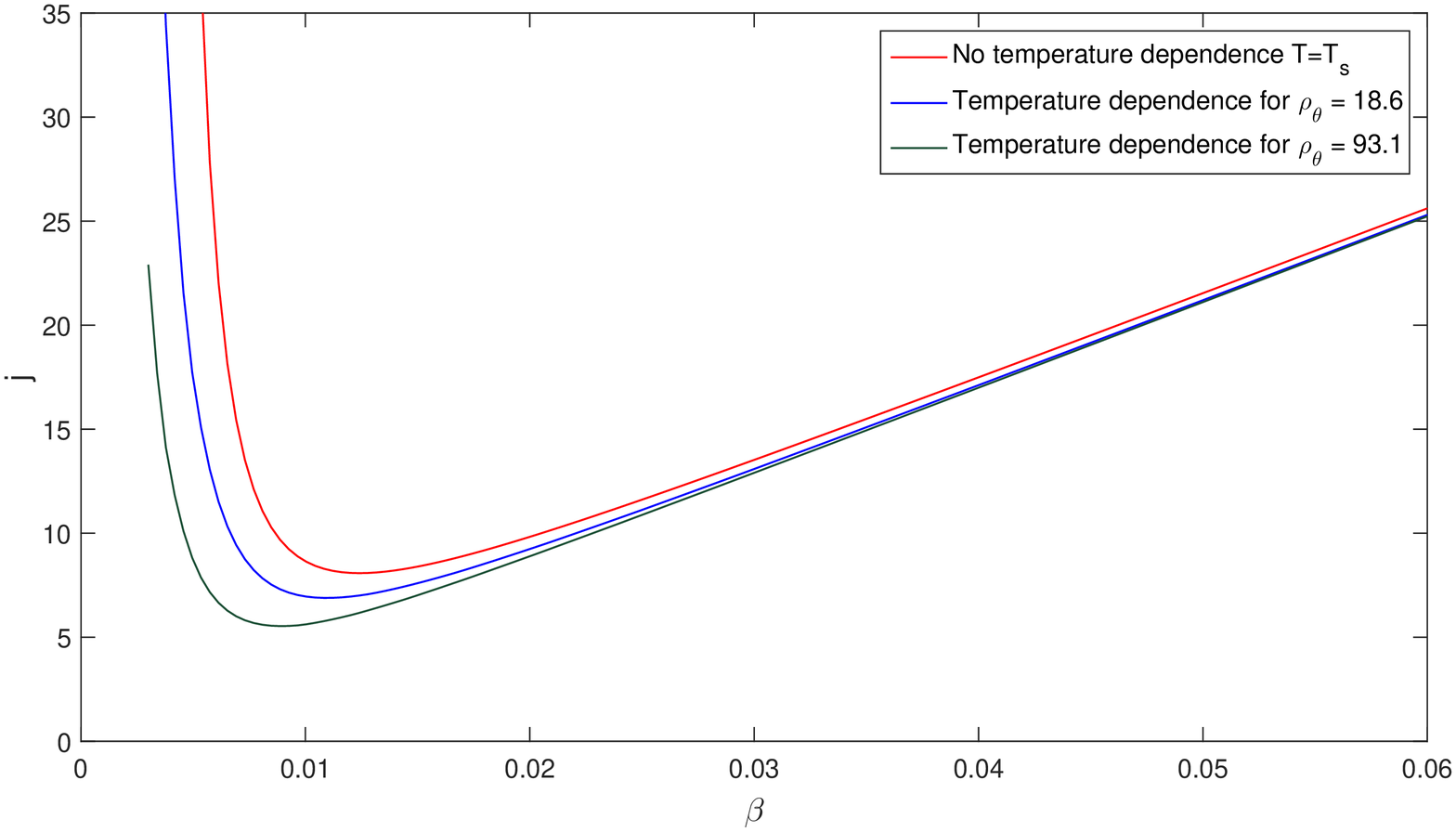}
\caption{Dimensionless current-voltage characteristics $j_\mathrm{_{2D}}(E)$ or rescaled $I(U)$ based on \Eqref{VAchar} and \Eqref{Kapitsa} for different interfacial thermal resistances.}
\label{KapitsaPlot}
\end{figure}

Since the fluctuational current formally tends to infinity at evanescent electric field bellow $T_c$, one can easily derive the asymptotics 
\begin{equation} 
\left. j(E) E\right \vert_{E \rightarrow 0} = (T_c - T_\mathrm{sub})/\mathcal{R}_\theta
=\left. \frac{1}{lw} U I(U) \right \vert_{U \rightarrow 0},
\label{ThermalRes}
\end{equation}
which allows us to determine the interface conductance $\mathcal{R}_\theta$.
Or else the current 
\begin{equation}
j(E)=\frac{T_c - T_\mathrm{sub}}{\mathcal{R}_\theta E} + \sigma_n E,
\end{equation}
meaning that we have hyperbolic approximation for the current-voltage characteristic.
Here the voltage $U$ and the current $I$ are given through the strip length $l$ and its width $w$. They are related to $j_\mathrm{_{2D}}$ and $E$ through
\begin{equation}
F(U) \equiv I=j_\mathrm{_{2D}} w, \quad U = E l.
\end{equation}
Below the critical temperature $T<T_c$ at small electric fields the current is significant and the temperature of the superconductor has to be calculated in a self-consistent way solving simultaneously \Eqref{Kapitsa} and \Eqref{VAchar} shown in \Fref{KapitsaPlot}.

The shape of the thin film is not necessarily a strip.
For example it can be in Corbino geometry, a narrow ring between radii $R_1$ and $R_2$ with $l=R_2 - R_1 << R_2$ and $w=2\pi R_1$.
Specifically for high $T_c$ superconductors, the frequency $1/\tau_0$ is in the terraherz range and for them the above-derived static formulas for the current are valid practically for the whole radio frequency range.
In the next section we present a circuit working within the framework of the theory presented so far.

\section{Circuit generating oscillations}

\begin{figure}[b]
\includegraphics[width=1\linewidth]{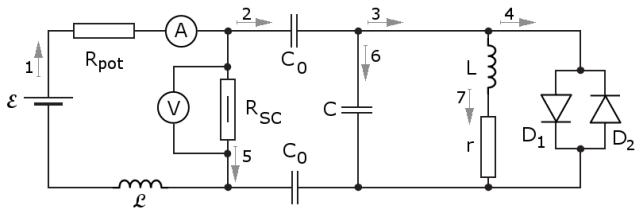}
\caption{Circuit with negative differential conductivity $R_\mathrm{SC}^{-1}$ of a superconductor supercooled below $T_c$ in an external electric field.
The direct current is created by a battery $\mathcal{E}$ and the current $I_1$ (denoted by the arrow with number 1 in the figure) measured by the ammeter A is regulated by the potentiometer $R_\mathrm{pot}$.
High-frequency current through the battery is stopped by the large inductance $\mathcal{L}$ and the voltmeter V measures the dc voltage of the thin superconductor film.
The dc current supply is galvanically separated from the parallel $LC$ resonator by the large capacitors $C_0$.
The resonance frequency $\omega_0=1/\sqrt{LC}$ is fixed by the values of the small capacitor $C$ and  the small inductance $L$ with internal resistance $r$.
The amplitude of the electric oscillations is limited by diodes $D_1$ and $D_2$.
The different currents in the circuit denoted in the figure by their numbers only, participate in the circuit equations \Eqref{eqI}-\Eqref{eqF}.}
\label{circuit}
\end{figure}

The circuit for generation of stable oscillations by using negative differential conductivity in supercooled superconductors is shown in \Fref{circuit}.
One can easily see the parallel $LC$ resonance circuit generating the oscillations.
In short, the negative differential resistivity $R_\mathrm{SC}$ amplifies the generated oscillations, while the two oppositely connected diodes $D_1$ and $D_2$ limit the amplitude of the oscillations.

Applying Kirchoff's laws, the dynamical laws governing the currents and voltages, to the circuit in \Fref{circuit}, the current and voltage equations are easily found to be
\begin{eqnarray}
\label{eqI}
\mathcal{E}-R_{\mathrm{pot}}I_1-U-\mathcal{L}\dot{I_1} &=& 0, \\
\dot{U}-\frac{2}{C_0} I_2-\frac{I_6}{C} &=& 0, \\ \label{Induct}
\frac{I_6}{C}-r \dot{I_7}- L \ddot{I_7} &=& 0, \\ 
I_5 - F(U) &=& 0, \\
I_1 - I_2 - I_5 &=& 0, \\
I_2 - I_3 - I_6 &=& 0, \\
I_3 - I_4 - I_7 &=& 0.
\label{eqF}
\end{eqnarray}
This system of equations governs the circuit operation.

The current through the diodes $D_1$ and $D_2$ ($I_4$) is governed by the volt-ampere characteristic
$$
I_4=\sigma_d\left[ (\mathcal{E}_L\!-\!U_0)\theta(\mathcal{E}_L\!-\!U_0)+(\mathcal{E}_L\!+\!U_0)(1\!-\!\theta(\mathcal{E}_L\!+\!U_0))\right],
$$
where $\mathcal{E}_L$ is the voltage between the two ends of the inductance $L$, $\sigma_d$ is the diode's conductance when it is "open" and $U_0$ is its opening voltage.
Obviously $I_4$ is entirely determined by the inductance's voltage, which from \Eqref{Induct} is completely determined by $I_7$ and $\dot{I_7}$. 

Eliminating $I_2$, $I_3$, $I_5$ and $I_6$ and defining the additional functions $Y(t)=\dot{I_7}(t)$
and
$$\mathcal{D}(x)= \sigma_d\left[ (x-U_0)\theta(x-U_0)+(x+U_0)(1-\theta(x+U_0))\right]$$
we can rewrite this system as
\begin{eqnarray*}
\dot{I_1} &=& \frac{\mathcal{E}}{\mathcal{L}}-\frac{R_{\mathrm{pot}}}{\mathcal{L}}I_1-\frac{U}{\mathcal{L}}, \\
\dot{I_7} &=& Y, \\
\dot{U} &=& \left( \frac{2}{C_0}+\frac{1}{C}\right)\left(I_1-F(U)\right) - \frac{1}{C}(\mathcal{D}(r I_7 + L Y)+ I_7), \\
\dot{Y} &=& \frac{1}{LC}(I_1 - I_7 - F(U)- \mathcal{D}(r I_7 + L Y)) - \frac{r}{L}Y. \\
\end{eqnarray*}
Defining the matrices $\psi(t)$, $\mathbf{N}(U)$ and $\hat{L}$ by
\begin{eqnarray}
\psi(t) &=& 
\begin{pmatrix}
I_1(t) \\
I_7(t) \\
U(t) \\
Y(t)
\end{pmatrix}, \nonumber \\
 \mathbf{N}(U,I_7,Y) &=&
 \begin{pmatrix}
\mathcal{E}/\mathcal{L} \\
0 \\
- \left( \frac{2}{C_0}\!+\!\frac{1}{C}\right) F(U)\!-\!\frac{1}{C}\mathcal{D}(r I_7\!-\!L Y) \\
-\frac{1}{LC}( F(U)+ \mathcal{D}(r I_7 + L Y)
 \end{pmatrix}, \nonumber \\
\hat{L} &=&
\begin{pmatrix}
-\frac{R_{\text{pot}}}{\mathcal{L}} & 0 & -\frac{1}{\mathcal{L}} & 0 \\
0 & 0 & 0 & 1 \\
 \left( \frac{2}{C_0}+\frac{1}{C}\right) & -\frac{1}{C} & 0 & 0 \\
 \frac{1}{LC} & -\frac{1}{LC}  & 0 & -\frac{r}{L} 
 \end{pmatrix}, \nonumber
\end{eqnarray}
we can write the system as the simple matrix equation
\begin{eqnarray}
\frac{\mathrm{d} }{\mathrm{d} t} \psi=\hat{L} \psi+ \mathbf{N}(U).
\end{eqnarray}

\section{Results}

\begin{figure}[h]
\includegraphics[width=1\linewidth]{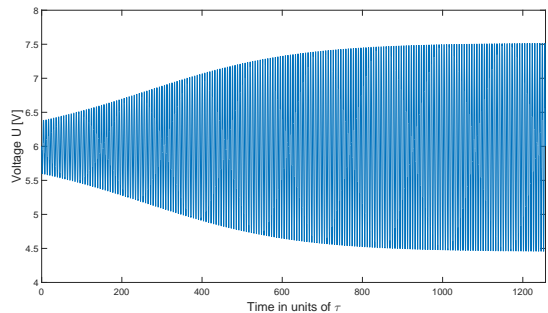}
\caption{Stable electric oscillations generated by the circuit in 
Fig.~\ref{circuit} using the values in Fig.~\ref{values}.
Heating of the layer stops the increment of the voltage amplitude.
The final amplitude is determined bi the interface thermal resistance between the superconductor layer and the substrate.
The abscissa is in dimensionless time in units $\tau=t/\sqrt{LC}$, i.e. for our example the time unit is 10~ns.}
\label{osc}
\end{figure}

\begin{figure}[h]
\includegraphics[width=1\linewidth]{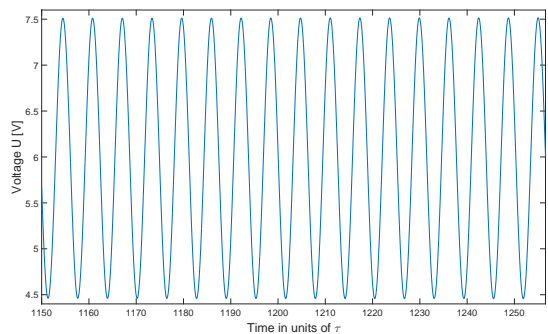}
\caption{Magnified region of the stable electric oscillations shown in \Fref{osc}.}
\label{osc_det}
\end{figure}

The calculated oscillations generated by the thin supercooled below $T_c$ superconductor from the circuit in Fig.~\ref{circuit} are shown in Fig.~\ref{osc} and magnified in Fig.~\ref{osc_det} with the numerical values presented in Table~\ref{values}.
\begin{table}[h]
\centering
	\caption{Numerical values of the circuit elements and parameters from \Fref{circuit} and the equations \Eqref{eqI}-\Eqref{eqF}.}
	\label{tbl:values}
\begin{tabular}{c  r c r}
		\hline
		Element  & Value & Parameter & Value \\ \hline
			$\mathcal{E}$ & 6~V & $l$ & 20~$\mu$m \\
			$R_\mathrm{pot}$ & 10~$\Omega$ & $w$ &  20~$\mu$m \\
			$\mathcal{L}$ & 1~mH  & $\xi$ & 1.5~nm\\ 
			$C_0$ & 1~mF & $S_n$ & 137 \\
			$C$ &   $1$~nF &  $T_\mathrm{sub}$ & 80~K \\ 
			$L$ &  100~nH &   $T_c$ & 90~K \\
			$r$ &  0.3~$\Omega$ & $\mathcal{R}_{\theta}$ & $6\times 10^{-14}$ $\mathrm{K \, m^2 \, W^{-1}}$ \\
			$U_0$ & 2~V &  & \\
			$\sigma_d$ &  0.1~Sm &  &  \\ 
\hline
\end{tabular}
\label{values}
\end{table}
The oscillations are in the GHz region as it is expected from the values of $L$ and $C$ from Table~\ref{values}.
The diodes are only a schematic element, they can be even omitted since the heating of the superconductor through the interface resistance $\mathcal{R}_\theta$ reliably limits the oscillations.
Non-linear effects associated with the boundary resistance in fact limit the oscillations amplitude preventing exponential growth.
The average dissipated power $ \langle j(E) E \rangle $ is less than 1~Watt.
In order to avoid boiling the nitrogen, the use of cryogenic electron microscopy may be required.
These results give an optimistic estimate that it is worth starting an experimental research with the available nanostructured samples -- a strip of thin film or 2-dimensional superconductor with 2 electrodes at the ends like source-drain channel in a field-effect transistor.

\section{Conclusions}

We demonstrate gigahertz range of electric oscillations created by the negative differential conductivity  in external electric field of supercooled below $T_c$ superconductor.
Using distributed elements, the frequency of generations is fixed but using as a resonator 2D plasmons,~\cite{Groshev} the resonance frequency can easily reach terahertz range.
The theory of damping rate of 2D plasmon will be given elsewhere.

The development work should start with investigation of current voltage characteristics of a nanostructured superconductor similar to a field effect transistor.
The area of source and drain should be maximal in order to ensure high currents.
If the gate is implemented by de-pairing magnetic material, the described sample will be a new type superconducting field effect transistor operating in the terahertz range.
Crucial starting point will be observation of annulation of differential conductivity at decaying electric field below $T_c$.
It was suggested by Gor'kov,~\cite{Gorkov:70} which tried to interpret the generation of high-frequency electric oscillations in thin superconducting films observed half a century ago.~\cite{Churilov:69}
Now the time has come to develop technical applications based on hybrid nanostructured superconductors.
For true 2D superconductors the temperature can be accepted to be equal to the ambient while for thin films it is indispensable to take into account the interface thermal boundary resistance.

\acknowledgments{}

Preliminary results from this research were presented in the COST School on Quantum Materials and Workshop on Vortex Behavior in Unconventional Superconductors, 7-12 October 2018 in Braga, Portugal and in the 10$^\mathrm{th}$ Anniversary International Conference on Nanomaterials -- R\&A, 17-19 October 2018 in Brno, Czech Republic.\cite{NANOCON:18}
Authors are thankful to Anna Palau, Andrey Varlamov and Hermann Suderow for the stimulating discussions during the COST school on quantum materials.

Authors will greatly appreciate any correspondence related to an experimental realisation of the proposed generator.

\end{document}